# Giant Modal Gain, Amplified Surface Plasmon Polariton Propagation, and Slowing Down of Energy Velocity in a Metal-Semiconductor-Metal Structure

D.B. Li and C.Z. Ning[1]

Department of Electrical, Computer and Energy Engineering and Center of Nanophotonics,

Arizona State University, Tempe, AZ 85287

[1]Corresponding author, email: cning@asu.edu

## Abstract

We investigated surface plasmon polariton (SPP) propagation in a metal-semiconductor-metal structure where semiconductor is highly excited to have optical gain. We show that near the SPP resonance, the imaginary part of the propagation wavevector changes from positive to hugely negative, corresponding to an amplified SPP propagation. The SPP experiences a giant gain that is 1000 times of material gain in the excited semiconductor. We show that such a giant gain is related to the slowing down of average energy propagation in the structure.

Surface plasmon polariton (SPP) is a coupled mode between an electronic excitation in metals (plasmons) and photonic mode at a metal-dielectric interface. While the basic phenomenon has been well-known for many decades, SPPs have attracted a great deal of attention recently due to interests ranging from fundamental physics to device applications [1]. This is especially true when the passive dielectric material is replaced by an active light emitting material such as dye molecules or semiconductor [2]. In a metal-semiconductor structure, the close proximity between excitons in semiconductor and plasmons in metal provide an interesting new setup to study the fundamental interactions between these two important elementary excitations mediated by photons of an extremely enhanced field. In terms of fundamental understanding of light-matter interactions, metal-semiconductor structures provide a unique platform where certain important issues can be reexamined such as validity of dipole approximation, strong field limit, and inhomogeneity at extremely small scales. From the device application point of view, SPPs in a metal-semiconductor structure provide probably the only likelihood of confining light field to a substantially small volume compared to wavelength, making active nanophotonics devices such as nanolasers possible.

Metallic structures are known to be able to guide or confine optical waves in a much smaller space dimension than the wavelength involved or the diffraction limit, thus becoming the natural choice for making nanophotonic devices such as waveguides and lasers at subwavelength scales or for studying light-matter interaction at a subwavelength scale. The detrimental loss in metals has been the only essential roadblock, leading to diminishing propagation length in a waveguide or insurmountable threshold gain for a nanolaser. The associated large loss hinders both experimental efforts in understanding fundamental physics and development of fundamentally new devices. Thus it is quite natural to consider integrating metals with semiconductor gain medium to compensate the metal loss. Among various structures that support SPP modes, metal-semiconductor-metal (MSM) structure is the canonical example and has attracted a great deal of attention recently for applications in active nanophotonic devices [3]-[4], where the adjacent semiconductor gain medium could compensate metal loss. Potential lossless propagation of surface plasmon polariton modes [3] and lasing [4]-[6] has been studied in these structures. It was shown [5] that a net modal gain is possible in a semiconductor-metal core-shell structure in a frequency band above the cut-off frequency, despite the large loss in the metal shell. The experimental demonstration of lasing in similar structure [6] partially verified the existence of this net gain. While existence of any net gain in a metal-semiconductor structure is important, the wavelength range near and above cut-off frequency and well-below SPP resonance corresponds to a greatly increased effective wavelength, since the propagation wavevector approaches zero near cut-off. Thus a key promise of plasmonic devices, the wavelength reduction, is not possible in this wavelength range. Since reduction of effective wavelength is the physical basis for size reduction of plasmonic devices, the best case scenario would be to achieve a net gain near the SPP resonance where the wavelength compression is the maximum. In that scenario, one would be able to achieve maximum reduction of wavelength (and device size) and positive gain simultaneously, leading to the smallest active devices. However, so far no theory or experiment has shown significant optical gain that can significantly over-compensate the metal loss near the SPP resonance. Here we demonstrate that, not only a modal gain exists in the MSM structure near the SPP resonance, but the gain is hugely enhanced. We discovered that the modal gain can be as large as 1000 times the semiconductor material gain. Furthermore, we show that this giant gain is associated with a significant slow-down of the average energy velocity in the structure.

We investigated the MSM plasmonic waveguide as our model system with special attention paid to the possibility of overall optical gain for the guided modes near the SPP resonance. The geometry of the MSM waveguide is shown in Fig. 1. A thin semiconductor core layer is sandwiched between two thick enough (usually larger than 100 nm) metal layers. The metal is silver with dielectric function $\varepsilon_2(\omega)$ [7], and the semiconductor has a dielectric constant $\varepsilon_1$ with a real part of 12. The material gain in the semiconductor can be modeled by adding a small negative imaginary part $\varepsilon_1''$ to the dielectric constant, and the material gain is given by $G_0 = -\varepsilon_1''\omega/(nc)$, where $n$ is the real part of refractive index. For the symmetric TM modes, the field components in different layers can be written as

| $x \le -d/2$ | $\lvert x \rvert \le d/2$ | $x \ge d/2$ |
|---|---|---|
| $H_y = A e^{k_2 x + i k_z z}$ | $H_y = B(e^{-k_1 x} + e^{k_1 x}) e^{i k_z z}$ | $H_y = A e^{-k_2 x + i k_z z}$ |
| $E_x = A \dfrac{k_z}{\omega \varepsilon_2 \varepsilon_0} e^{k_2 x + i k_z z}$ | $E_x = B \dfrac{k_z}{\omega \varepsilon_1 \varepsilon_0} (e^{-k_1 x} + e^{k_1 x}) e^{i k_z z}$ | $E_x = A \dfrac{k_z}{\omega \varepsilon_2 \varepsilon_0} e^{-k_2 x + i k_z z}$ |
| $E_z = -A \dfrac{i k_2}{\omega \varepsilon_2 \varepsilon_0} e^{k_2 x + i k_z z}$ | $E_z = B \dfrac{i k_1}{\omega \varepsilon_1 \varepsilon_0} (e^{-k_1 x} - e^{k_1 x}) e^{i k_z z}$ | $E_z = A \dfrac{i k_2}{\omega \varepsilon_2 \varepsilon_0} e^{-k_2 x + i k_z z}$ |

where $A/B = e^{-(k_1 - k_2) d/2} + e^{(k_1 + k_2) d/2}$. The dispersion relation is determined by

$$\varepsilon_2 k_1 \tanh(k_1 d/2) = -\varepsilon_1 k_2 , \tag{1}$$

and the propagation wavevector is defined by

$$k_z^2 = \varepsilon_{1,2} k_0^2 + k_{1,2}^2 . \tag{2}$$

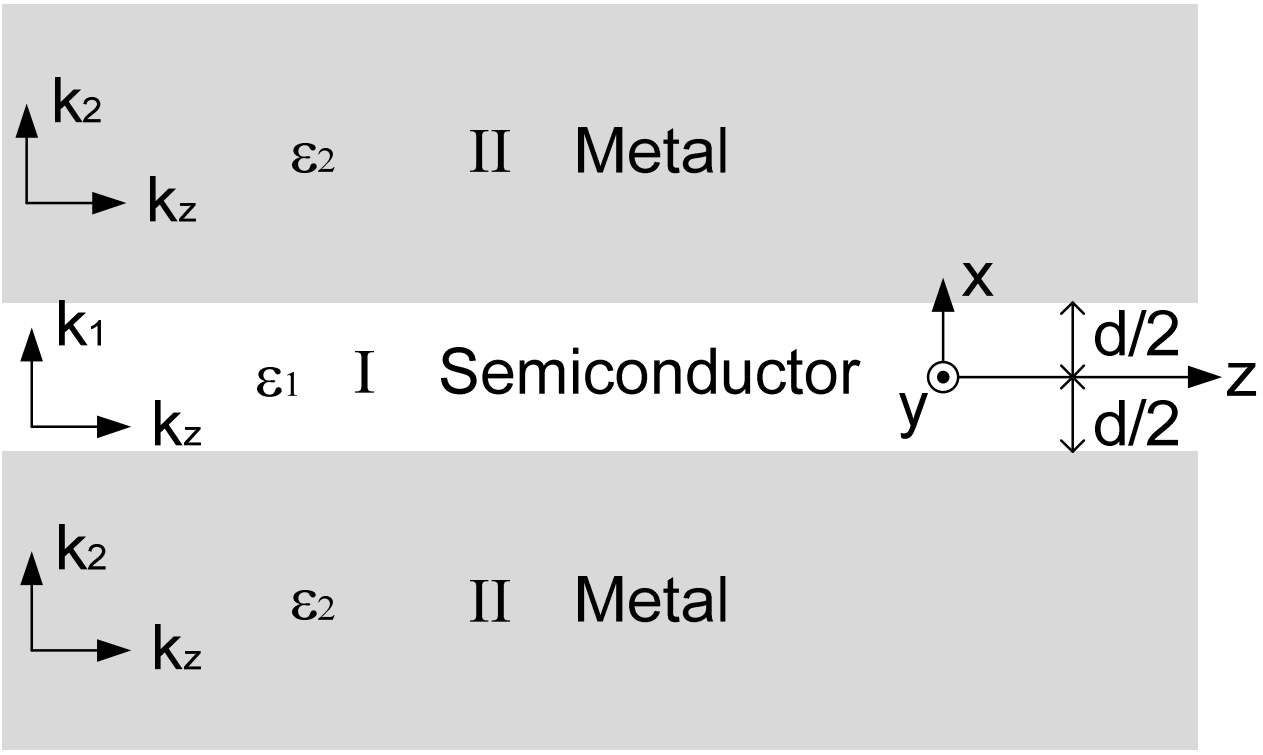

FIG. 1 Geometry of a MSM waveguide and our notations. The dielectric constants of semiconductor and metal are $\varepsilon_1$ and $\varepsilon_2$ respectively. The thickness of semiconductor layer is *d*.

It is easy to show that the eigenvalue equation (1) for the MSM structure becomes the well-known expression for the SPP mode along a MS interface in the limit of large $|k_1 d|$, ( $\tanh(k_1 d/2) \approx 1$ ): $\varepsilon_2 k_1 \approx -\varepsilon_1 k_2$ . In a more general case of finite thickness, *d*, eigenvalue equation (1) has to be solved numerically.

Figure 2(a) and 2(b) show respectively the real and imaginary parts of $k_z$ as a function of photon energy with *d*=100nm and 200nm for 4 different values of $\varepsilon_1''$, compared with the dispersion relations in a MS bi-layer structure. It is seen that all structures have the same behavior around SPP resonance (within a range of $\pm 25$ meV around the peak), independent of the number of interfaces or the thickness of the middle layer. This means the guided modes in the MSM waveguide are decoupled into two independent SPP modes at the two MS interfaces. We observe in Fig. 2(a) that optical gain in semiconductor layer can significantly modify the resonance behavior [8], as can be easily seen by comparing different panels of Fig. 2(a). There exists an optimum level of semiconductor gain (around $\varepsilon_1'' = -0.4$ ), where SPP resonance has the maximum response (with highest and narrowest resonance peak). For application of active optical devices, it is more important to study the imaginary part of $k_z$ as it describes the optical loss or gain. The *modal* gain is given by, $G_m$ ( $=-2\text{Im}[k_z]$ ), which describes the overall gain that a given mode experiences in a waveguide containing active region, while the material gain, $G_0$ defined earlier describes how a plane wave is amplified in an infinitely large medium. For example, the intensity of a electromagnetic wave traveling along the *z*-axis of the waveguide can be written as $I = I_0 e^{G_m z}$, where $I_0$ is the initial intensity. In Fig. 2(b) we see that, for small material gain (e.g., when $\varepsilon_1'' > -0.3$ ), Im[$k_z$] is positive or modal gain is negative near SPP resonance due to metal loss. However, Im[$k_z$] becomes extremely negative or modal gain becomes a giant positive value near the SPP resonance when the semiconductor material gain $G_0$ is large enough (or $\varepsilon_1''$ is enough negative). For $\varepsilon_1'' = -0.4$, which corresponds to a material gain of about $1.35\times10^4$ cm$^{-1}$, we see that the peak modal gain becomes much larger than the material gain. It may sound counter-intuitive that a guided mode can experience more gain (modal gain) per unit propagation length than the gain provided by the gain medium (material gain). The ratio of the two is known as confinement factor in conventional waveguide with gain material. This was indeed an issue of debate when a larger-than-unity confinement factor was first shown in the case of a semiconductor nanowire [9]. In that case, the group velocity (defined as $v_g = \partial\omega / \partial(\mathrm{Re}[k_z])$) slows down dramatically in a strongly guided situation, leading to a large optical gain per unit length.

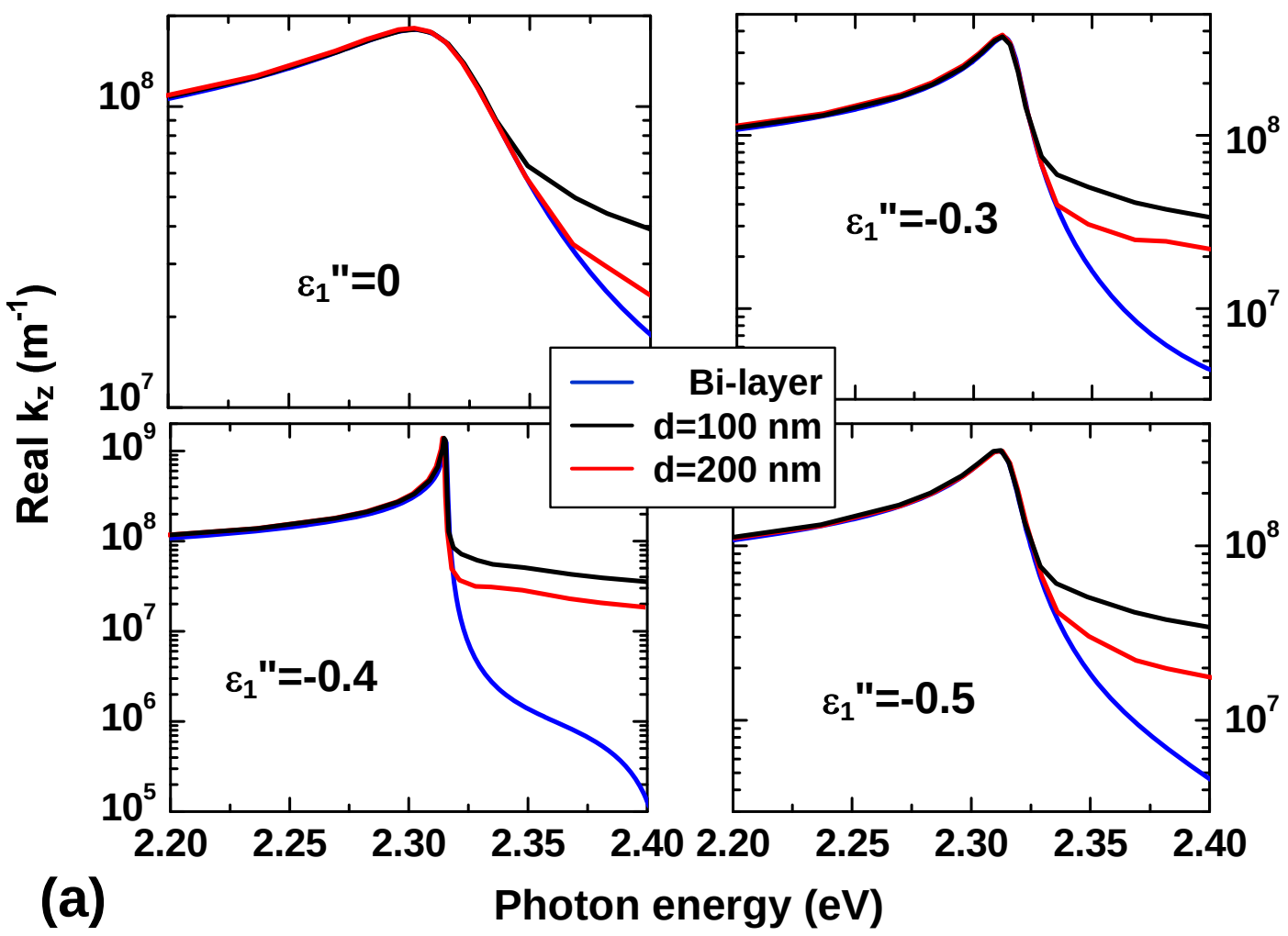

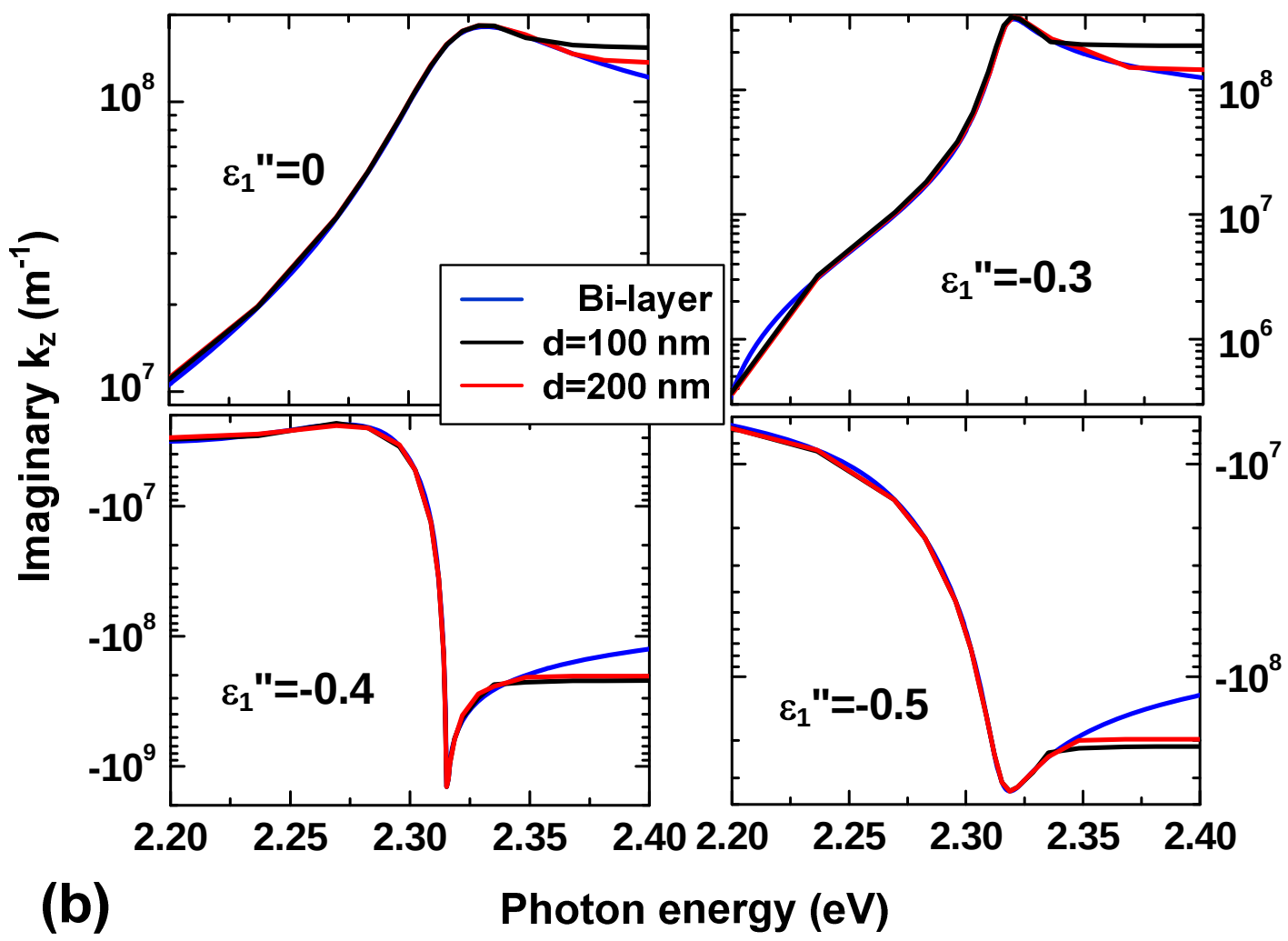

FIG. 2 Real and imaginary parts of wavevector $k_z$. (a) The real part and (b) the imaginary part of $k_z$ as a function of photon energy with $d$=100nm (black) and 200nm (red) for 4 different values of $\varepsilon_1''$. The dispersion relations in a MS bi-layer structure (blue) are also shown for comparison.

To see if the huge enhancement of modal gain is related to the slowing down of group velocity, we plot in Fig. 3(a) the group velocity for $\varepsilon_1'' = -0.4, -0.5$, compared with the corresponding modal gain in Fig. 3(b). As we expect, the group velocity in a plasmonic waveguide can take both positive and negative values around the plasmon resonance. However, the maximum of modal gain does not correspond to the minimum of the group velocity. To gain further understanding of this issue, we recall that, in the dispersive medium with loss or gain, a physically more fundamental velocity is not group velocity, but energy velocity $v_E$, or the velocity of energy propagation for a particular mode. Energy velocity [10] is defined as the ratio of energy flux density $S$ over stored energy density $w$, i.e., $v_E=S/w$. Since energy propagates in opposite directions in semiconductor and metal, an average energy velocity $\overline{v}_E$ is introduced and is given by:

$$\bar{v}_E = \frac{\int w^s v_E^s d\mathbf{v} + \int w^m v_E^m d\mathbf{v}}{\int w^s d\mathbf{v} + \int w^m d\mathbf{v}} = \frac{\int S^s d\mathbf{v} + \int S^m d\mathbf{v}}{\int w^s d\mathbf{v} + \int w^m d\mathbf{v}} \tag{3}$$

where superscripts "*s*" and "*m*" represent quantities in semiconductor and metal respectively. The *z*-component of the energy flux density $S_z$ in this case can be written as

$$S_z = \frac{1}{2}\mathrm{Re}[E_x H_y^*], \tag{4}$$

and the energy density in semiconductor is defined in the usual way as

$$w^s = \frac{1}{2}\mathrm{Re}[\mathbf{E}^s \cdot \mathbf{D}^{s*} + \mathbf{B}^s \cdot \mathbf{H}^{s*}], \tag{5}$$

while energy density in metal is derived from Poynting's theorem in linear, lossy, and dispersive media [10]:

$$w^m = \frac{1}{2}\mathrm{Re}[\frac{d(\omega\varepsilon_2)}{d\omega}\mathbf{E}^m \cdot \mathbf{E}^{m*} + \mathbf{B}^m \cdot \mathbf{H}^{m*}]. \tag{6}$$

Figure 3(c) shows the z-component of the average energy velocity as a function of photon energy for $\varepsilon_1'' = -0.4, -0.5$. It is seen that $\bar{v}_E$ has its minimum value whenever $G_m$ is the largest. This relationship is true for all the values of the material gain. This means that the giant modal gain comes from the slowing down of the average energy propagation. A slowed energy transport allows more energy exchange events between waves and gain/loss media, or a slow wave will experience more absorption or emission events than fast wave when traveling through the same distance. If the total system is lossy, this enhanced exchange will lead to large modal loss near resonance; in the opposite situation, it will lead to a huge modal gain. As can be seen from the figure, a modal gain as much as $2\times10^7$ cm$^{-1}$ is achieved near the resonance. This represents a modal gain as large as 1000 times of the material gain ($G_0 \approx 1.35\times10^4$ cm$^{-1}$ for $\varepsilon_1'' = -0.4$) in the semiconductor.

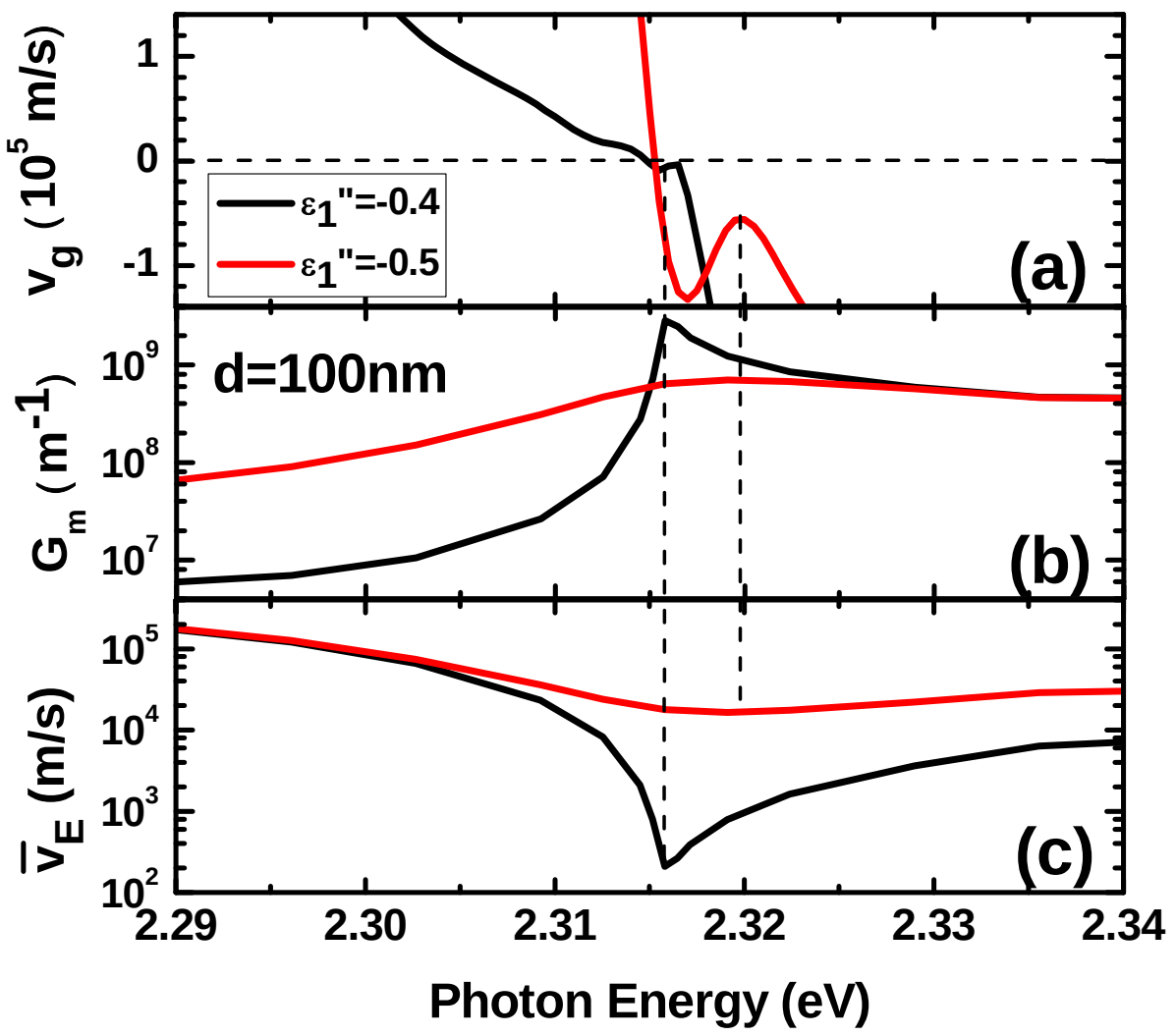

FIG. 3 Comparison of energy velocity and modal gain (a) group velocity, (b) modal gain and (c) average energy velocity along *z*-axis as a function of photon energy for $\varepsilon_1'' = -0.4$ (black), −0.5 (red). The thickness of the semiconductor layer is 100nm.

While the enhancement of modal gain is obviously of great significance, this only occurs when the material gain is sufficiently large to balance out the metal loss and to achieve such a huge modal gain. It is thus important to assess the feasibility of achieving such material gain. For example, in order to obtain the giant modal gain in the MSM waveguide with 100 nm-thick middle layer, the required minimum material gain ($G_{min}$) is about $1.34\times10^4$ cm$^{-1}$ (corresponding to $\varepsilon_1'' \approx -0.396$). While material gain of this magnitude or as large as $2\times10^4$ cm$^{-1}$ has been shown possible in widegap semiconductor quantum wells such as nitrides or II-VI semiconductors [11], it is more desirable to have this gain enhancement occur at a lower material gain level. A possible initial experiment can be performed at low temperature. As is well known, optical loss in metals such as silver is significantly reduced at low temperature, while gain in semiconductor is significantly increased. It was shown experimentally that the imaginary part of silver dielectric constant is reduced as temperature decreases [12]. The reduction rate at 3 eV is about $5\times10^{-4}$ $K^{-1}$. Rogier *et al.* [13] theoretically estimated the temperature coefficients of both real and imaginary parts of silver dielectric constant at 2 eV to be $8.5\times10^{-4}$ $K^{-1}$ and $1.5\times10^{-3}$ $K^{-1}$ respectively. Using the experimental results in Ref. [12], we estimate that the required imaginary part of the semiconductor dielectric constant can be as low as -0.271 at 77 K to achieve giant modal gain, corresponding to a material gain ($G_{\min}$) of $9\times10^{3}$ $cm^{-1}$. To be more specific, let's consider a MSM waveguide with $Zn_{0.8}Cd_{0.2}Se$ as the middle layer material at 77 K. The material gain of $Zn_{0.8}Cd_{0.2}Se$ was calculated using a free-carrier model [14], using material parameters of ZnSe and CdSe in Ref. [15]. The material parameters of $Zn_{0.8}Cd_{0.2}Se$ were obtained by linear interpolation. The calculated gain $G_0$ spectrum is shown in Fig. 4 (dashed line) with carrier density $1.1\times10^{19}$ $cm^{-3}$. Such gain level at 77 K is reasonable in $Zn_{0.8}Cd_{0.2}Se$ [16]. The actual spectral dependence of imaginary part of the dielectric function obtained using $\varepsilon_1'' = -ncG_0 / \omega$ is then used to solve for the propagation wavevector, $k_z$, which is shown for the fundamental TM mode in Fig. 4 (solid lines). As we see, the similar huge negative imaginary $k_z$ (red curve) occurs near SPP resonance, which means the giant modal gain occurs in this case. In addition, this specific example also justifies our earlier treatment of imaginary part of semiconductor as a frequency independent value, since semiconductor gain is much broader than the imaginary part of the wavevector near SPP resonance and similar results are obtained using the realistic spectral dependence.

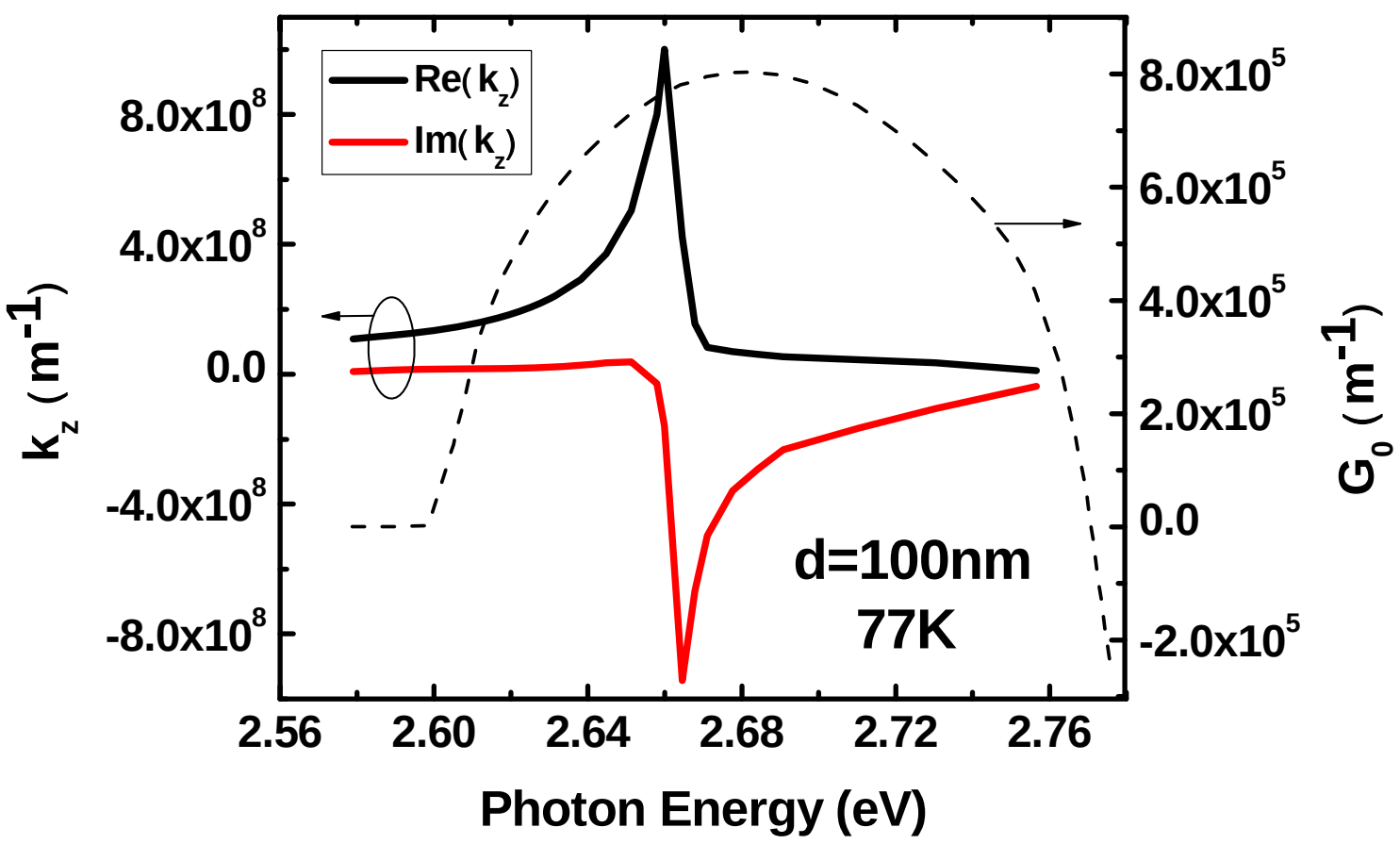

FIG. 4 Dispersion curves in a silver/$Zn_{0.8}Cd_{0.2}Se$/silver waveguide with *d*=100 nm at 77 K. Real (solid black line) and imaginary (solid red line) part of $k_z$ was calculated for the fundamental TM mode. The material gain $G_0$ spectrum (dashed line) in the $Zn_{0.8}Cd_{0.2}Se$ layer with carrier density $1.1\times10^{19}$ $cm^{-3}$ is also shown.

To summarize, we have discovered a giant modal gain in a metal-semiconductor-metal plasmonic waveguide near the SPP resonance. We showed that the giant modal gain is related to the slowing down of average energy velocity, rather than the group velocity as in dielectric waveguide. This realization will contribute significantly to our rapidly increasing understanding of SPP phenomena in active plasmonics and will likely to stimulate new experiments in the field. Such unprecedented modal gain is also expected to have great impact to many plasmonic devices such as plasmonic waveguides with significantly increased propagation length and to sub-wavelength plasmonic nanolasers or surface plasmon polariton lasers.

Acknowledgements

This work was supported by Defense Advanced Research Project Agency (DARPA)'s NACHOS program.

[1] J. Seidel, S. Grafström, and L. Eng, Phys. Rev. Lett. **94**, 177401 (2005); Y. Fedutik *et al*., Phys. Rev. Lett. **99**, 136802 (2007).

[2] N. W. Lawandy, Appl. Phys. Lett. **85**, 5040 (2004); M. A. Noginov *et al*., Opt. Express **16**, 1385 (2008); M. Ambati *et al*., Nano Lett. **8**, 3998 (2008); M. A. Noginov *et al*., Phys. Rev. Lett. **101**, 226806 (2008).

[3] M. P. Nezhad, K. Tetz, and Y. Fainman, Opt. Express **12**, 4072 (2004); S. A. Maier, Opt. Commun. **258**, 295 (2006).

[4] T. Okamoto, F. H'Dhili, and S. Kawata, Appl. Phys. Lett. **85**, 3968 (2004); G. Winter, S. Wedge, and W. L. Barnes, New J. Phys. **8** 125 (2006).

[5] A. V. Maslov and C. Z. Ning, in Proc. of SPIE on Physics and Simulation of Optoelectronic Devices XV, edited by M. Osinski *et al*., **6468**, 64680I (2007)

[6] M. T. Hill *et al*., Nat. Photonics **1**, 589 (2007).

[7] P. B. Johnson and R. W. Christy, Phys. Rev. B **6**, 4370 (1972).

[8] I. Avrutsky, Phys. Rev. B **70**, 155416 (2004); M. A. Noginov *et al*., Opt. Lett. **31**, 3022 (2006).

[9] A. V. Maslov and C. Z. Ning, J. Quantum. Electron. **40**, 1389 (2004).

[10] L. D. Landau and E. M. Lifshitz, *Electrodynamics of Continuous Media*, 2nd. edn (Butterworth-Heinenann, Oxford, 1984).

[11] W. W. Chow and S. W. Koch, *Semiconductor-Laser Fundamentals: Physics of the Gain Materials* (Springer-Verlag, Berlin Heidelberg, 1999).

[12] P. Winsemius *et al.*, J. Phys. F: Metal Phys. **6**, 1583 (1976).

[13] R. H. M. Groeneveld, R. Sprik and A. Lagendijk, Phys. Rev. Lett. **64**, 784 (1990).

[14] W.W. Chow and S.W. Koch, *Semiconductor-Laser Fundamentals,* (Springer Verlag, Berlin-Heidelberg, 1999).

[15] E. D. Palik, *Handbook of Optical Constants of Solids* (Academic, New York, 1985); M. C. Tamargo, *II-VI semiconductor materials and their applications*, edited by M. O. Manasreh, Optoelectronic Properties of Semiconductors and Superlattices, Vol. 12 (CRC, 2002).

[16] D. Ahn, T.-K. Yoo and H. Y. Lee, Appl. Phys. Lett. **59**, 2669 (1991); Y.-H. Wu, J. Quantum Electron. **30**, 1562 (1994).